\documentstyle[prl, aps, twocolumn, psfig]{revtex}

\newcommand\be{\begin{equation}}
\newcommand\ee{\end{equation}}
\newcommand\bd{\begin{displaymath}}
\newcommand\ed{\end{displaymath}}
\newcommand\bea{\begin{eqnarray}}
\newcommand\eea{\end{eqnarray}}

\begin{document}

\twocolumn[

\hsize\textwidth\columnwidth\hsize\csname @twocolumnfalse\endcsname

\draft

\title{Lattice-Independent Approach to Thermal Phase Mixing }

\author{Carmen J. Gagne and
Marcelo  Gleiser}
\address{Department of Physics and Astronomy,
Dartmouth College, Hanover, NH 03755, USA}

\maketitle

\begin{abstract}
We show how to achieve lattice-spacing independent results in numerical
simulations of finite-temperature stochastic scalar field theories. We
generalize the previous approach of Ref. \cite{BG}  by obtaining results
which are independent of the renormalization scale. As an application of
our method, we examine thermal phase mixing in the context of Ginzburg-Landau
models with short-range interactions.
In particular, we obtain the lattice-spacing and renormalization-scale
independent critical value of the control parameter which determines the
free-energy barrier between the two low-temperature phases. We also propose a
simple procedure to extract the critical value of control parameters for
different choices of lattice spacing.
\end{abstract}


]

\section{Introduction}

During the past decade, the study of equilibrium and nonequilibrium
dynamics of field theories has greatly benefited from the widespread
availability of workstations capable of millions of floating point
operations per CPU-second.

One of the most popular applications of computers in the physical
sciences is the examination of phenomena which are generated by
nonperturbative effects. These include nonlinear dynamical systems with a
few, several, or an infinite number of degrees of freedom. Of these, we
are particularly interested here in the latter, as they represent a
unique challenge to computational physics. Implementing field theories in
the computer implies discretizing not only time but also space: the
system is cast on
a finite lattice with a discrete spatial step,
effectively cutting off the theory both in the
infrared (by the lattice
size) and in the ultraviolet
(by the lattice spacing).  Although in classical
field theories an ultraviolet cutoff solves the Rayleigh-Jeans
ultraviolet catastrophe, the solution comes with a
high price tag:
whenever there is dynamical mixing of short and long wavelength modes,
the results will in general depend on the shortest distance scale in the
simulation, the lattice spacing. To be sure, in many
instances this dependence on small spatial scales does not affect
qualitatively the
physics one is interested in: for example, very near criticality for
Ising systems, where spatial correlations in
the order parameter diverge \cite{GOLDENFELD}, or are controllable in some
way \cite{PLASMA}. However, in many other cases one is interested in
achieving a proper continuum limit on the lattice which is independent of
the choice of ultraviolet cutoff. These include a wide range of phenomena
which have triggered much recent interest, from pattern formation in
fluid dynamics \cite{PATTERNS} to simulations of phase
transitions and topological defect formation, which often use stochastic
methods \cite{CPT}.

In the present work, we are concerned with curing, or at least greatly
alleviating, the lattice-spacing dependence that appears in stochastic
simulations of scalar field theories. We will show that it is indeed
possible to obtain results which are lattice-spacing independent, as long
as proper counterterms are added to the lattice effective potential.
Following a suggestion by Parisi \cite{PARISI}, lattice-spacing
independent results were recently obtained by J. Borrill and
one of us within the
context of finite-temperature symmetry restoration in a simple
Ginzburg-Landau model \cite{BG}. However, that study focused on a regime
where the large temperatures needed for symmetry restoration compromised
the approach to obtain lattice-spacing independent results, which is
based on a perturbative expansion in powers of the temperature.
Furthermore, no attempt was made to obtain results which were independent
of the renormalization scale. Thus, in that study, the numerical
prediction for the critical temperature depends on the particular choice
of renormalization scale.

Here, we would like to apply an expanded version of the method proposed
in Ref. \cite{BG} to a related problem, phase mixing in Ginzburg-Landau
models. The distinction between phase mixing and symmetry restoration is
made clearer through the following argument. Suppose a system
described by a Ginzburg-Landau free energy density with {\it odd} powers of
the order parameter $\phi (t,{\bf x})$ is rapidly cooled from a high
temperature to a temperature where two phases can coexist. The system was
cooled so as to remain entirely in one of the two phases. The odd term(s)
could be due to an external magnetic field (linear term), or to the
integration of other fields coupled to $\phi$, as in certain gauge
theories (cubic term) \cite{EWPOT}, or in de Gennes-Landau models of
the nematic-isotropic transition in liquid crystals \cite{LIQCRYS}. Due
to the odd terms, there is a free-energy barrier for large-amplitude
fluctuations between the two phases. [Of course, small-amplitude fluctuations
within each phase are also possible, but less interesting.]
This barrier is usually controlled by the
coefficients of the odd terms in the Ginzburg-Landau model, which we will
call the control parameter(s).

Suppose now that the system is held at the temperature where the two
phases have the same free energy densities (sometimes called the critical
temperature in the context of discontinuous phase transitions) and that
we are free to change the value of the control parameter(s). The question
we would like to address is how does the system behave as a function of
the control parameter(s), that is, as the free-energy barrier for
large-amplitude fluctuations between the two phases is varied. As is
well known, the mean-field theory approach breaks down when
large-amplitude fluctuations about equilibrium become large enough. Thus,
we should expect that the prediction from the Ginzburg-Landau model, that
the system remains localized in one phase until the barrier disappears
(when the control parameter goes to zero) will eventually be wrong. There
will be a critical value for the control parameter beyond which
nonperturbative effects lead to the mixing of the two phases. [Note that
due to the odd terms, there is no symmetry to be restored.] In the
language of the Ginzburg-Landau model, the system should at this point be
described as having a single well, centered at the mean value of the
order parameter. We would like to obtain the lattice-independent critical
value of the control parameter for this phase mixing to occur.

It is important to distinguish between phase coexistence and phase mixing. As
is well-known, phase coexistence will generally occur when a system is cooled
into the so-called phase coexistence region of the phase diagram. In this case,
the system will relax into its lowest free-energy configuration via spinodal
decomposition. Here, we are preparing the system initially outside the
phase coexistence region, namely in one particular phase only. In the
infinite-volume limit, mean field theory predicts the system will remain
there, since, as the two minima are degenerate, the nucleation of a critical
droplet would cost an infinite amount of free energy. Phase mixing is a
nonperturbative phenomenon characterized by large-amplitude fluctuations not
included in the mean-field approach. It signals the breakdown of mean-field
theory.

This paper is organized as follows:
In the next section we describe the
continuum model we use
and some of its properties. In section 3 we
describe the lattice implementation and how simulations using a bare
lattice potential give results which depend severely
on the lattice
spacing. In section 4 we show how to cure this dependence, and also how
to make the simulations independent of the choice of renormalization
scale. We conclude in section 5 with a brief summary of our results and a
discussion of future work.

\section{The Model in the Continuum}

Our starting point is the 2-dimensional Hamiltonian, (we use $c=k_B=1$)
\begin{equation}
{{H[\phi]}\over T} = {1\over T}\int d^2x\left [{1\over 2}\left
(\nabla\phi\cdot\nabla\phi\right )+
V(\phi) \right ]~~,
\label{hamiltonian}
\end{equation}
where the homogeneous part of the free energy density is
\begin{equation}
 V(\phi) = \frac{a}{2} (T^2-T^2_2) \phi^2 - \frac{\alpha}{3} T \phi^3 +
\frac{\lambda}{4} \phi^4 ~~.
\label{gl pot}
\end{equation}
This choice of $V(\phi)$ is inspired by several models of nucleation in
the condensed matter \cite{BINDERNUC} and high-energy physics literature,
in particular in recent models of the electroweak phase transition
\cite{EWPOT}. The several parameters in $V(\phi)$ allow one to apply it
to several situations of interest. However, we note that
here the order parameter is a scalar quantity, and thus the critical behavior
of this model belongs
to the universality class of the 2-dimensional Ising model
\cite{ISING,BG}. It is quite straight forward to generalize our results to
systems in different numbers of spatial dimensions.

At the critical temperature $T_c^2 = T_2^2/(1-2\alpha^2/9a\lambda)$, the
system exhibits two degenerate free energy minima at
\begin{equation}
\phi=0~~ {\rm and}~~ \phi_+={{2\alpha T_c}\over {3\lambda}}~~,
\label{tc minima}
\end{equation}
while at the temperature $T_2$ the barrier between the two phases
disappears. Throughout this work, we will be interested in the behavior
of the system at $T_c$. One reason for this choice has to do with the use
of a perturbative expansion which is in powers of $T$; at $T_c$ the expansion
parameter is sufficiently small, allowing us to stay at 1-loop. Another reason 
is
that we are interested in measuring the breakdown of mean-field theory in terms
of
parameters controlling the free-energy barrier, and the calculations are much
simpler at $T_c$, as we will see next.

According to the model described by Eq. \ref{gl pot},
at $T_c$, unless $\alpha=0$ (or $\lambda \to \infty $) there
will always be a barrier separating the two phases: at $T_c$, this
model does not
predict phase mixing to occur. It is thus very
convenient to introduce the shifted field
\begin{equation}
\phi \to \phi'\equiv \phi - \frac{T \alpha}{3 \lambda}~~,
\end{equation}
and write the shifted homogeneous free energy density as (dropping the
primes)
\begin{equation}
V_0(\phi) =
-\frac{1}{2}\mu^2(T)\phi^2+\frac{\lambda}{4}\phi^4+A(T)\phi+{\rm
constants}~~,
\label{shifted pot}
\end{equation}
with
\begin{equation}
\mu^2(T) \equiv -a(T^2-T^2_2) + \frac{T^2 \alpha^2}{3 \lambda}~~,
\end{equation}
and
\begin{equation}
A(T) \equiv a ( T^2 -T^2_2)\frac{T \alpha}{3 \lambda} +
\frac{2}{27}\frac{T^3 \alpha^3}{\lambda^2}~~.
\end{equation}
The shifted free energy density is just the usual Ginzburg-Landau free
energy density with an external magnetic field $A(T)$. Note that
$A(T_c)=0$ and the two minima are degenerate, as they should be. We now
introduce the dimensionless variables $\theta \equiv T/T_2$, $\tilde t
\equiv \sqrt{a} T_2 t$,
$\tilde x \equiv \sqrt{a} T_2 x$,
$\tilde\phi\equiv\frac{\phi}{\sqrt{T_2}}$, according to which we can
write, at
$\theta_c = \left[1-\frac{2 \tilde\alpha^2}{9 \tilde\lambda}
\right]^{-1/2}$,
\begin{equation}
\tilde V_0 = -{1\over 2}\tilde \mu^2(\theta_c)\tilde \phi^2 +{1\over
4}\tilde \lambda\tilde \phi^4~~,
\label{dimensionless pot}
\end{equation}
where $\tilde\lambda\equiv\frac{\lambda}{aT_2}$, $\tilde\alpha \equiv
\frac{\alpha}{a \sqrt{T_2}}$, and
\begin{equation}
\tilde\mu^2 = \frac{\mu^2}{a T_2} = -(\theta_c^2-1)+
                             \frac{\theta_c^2 \tilde\alpha^2}{3
\tilde\lambda}~~.
\end{equation}
Since we will keep the system at $\theta_c(\tilde \alpha,\tilde
\lambda)$, the only two control parameters are $\tilde\alpha$ and
$\tilde\lambda$. In what follows, we will fix $\tilde \lambda=0.1$ for
simplicity. This was also the choice in a previous study of phase mixing
in the same system, which did not address the issue of lattice-spacing
dependence \cite{GLEISER94}. We will also drop all tildes, except in the plots
and captions, where unshifted, dimensionless variables are used and marked 
explicitly.

\section{Numerical Results: Bare Lattice}

\subsection{Description of the Simulation}
As mentioned in the introduction, we would like to study the behavior of
the system described in the previous section when coupling to an external
thermal bath promotes fluctuations about equilibrium. We will consider
the situation where the system is initially prepared in the phase given
by $\phi=0$ in the unshifted potential or, more generically, the left
well. Since we are only interested in the final equilibrium value of the
system, we will simulate the coupling of the scalar field $\phi$ to the
thermal bath using a generalized Langevin equation,
\begin{equation}\label{e:langevin}
{\partial^2\phi\over\partial t^2} = \nabla^2\phi - \eta
{\partial\phi\over\partial t}
- {\partial V_0 \over \partial \phi} + \xi({\bf x},t)~~,
\end{equation}
where the viscosity coefficient $\eta$, set equal to unity in all simulations,
is related to the
stochastic force of zero mean $\xi({\bf x},t)$ by the
fluctuation-dissipation relation,
\begin{equation}\label{e:fluct-diss}
\langle \xi({\bf x}, t) \xi({\bf x}', t') \rangle = 2 \eta \theta 
\delta({\bf x} - {\bf x}') \delta(t - t')~.
\end{equation}

The system is discretized and put on a square lattice with side length, $L$,
equal to 64
for all the simulations,  but several lattice spacings,  $\delta x$, and
time steps,
$\delta t$, are used.  For $\delta x= $1.0, 0.8, and 0.2 the respective
time steps
are $\delta t=$ 0.2, 0.1, and 0.02. We have, of course, checked the stability of
the program for these choices of lattice parameters.
Using a standard second-order staggered leapfrog method (which is
second order in both space and time)
we can write,
\begin{eqnarray}\label{e:lattice_equation}
\dot{\phi}_{i,m+\frac{1}{2}} & = & \frac{(1 - \frac{1}{2} \eta \delta t)
\dot{\phi}_{i,m-\frac{1}{2}} + \delta t (\nabla^{2} \phi_{i,m} -
V'_0(\phi_{i,m}) + \xi_{i,m})}{1 + \frac{1}{2} \eta \delta t}
\nonumber \\
\phi_{i,m+1} & = & \phi_{i,m} + \delta t \dot{\phi}_{i,m+\frac{1}{2}}
\end{eqnarray}
where $i$-indices are spatial and $m$-indices temporal, overdots
represent derivatives with respect to $t$ and primes with
respect to $\phi$. The discretized fluctuation-dissipation
relation now reads
\begin{equation}
\langle \xi_{i,m} \xi_{j,n} \rangle = 2 \eta \theta 
\frac{\delta_{i,j}}{\delta x^{2}} \frac{\delta_{m,n}}{\delta t}
\end{equation}
so that
\begin{equation}
\xi_{i,m} = \sqrt{\frac{2 \eta \theta}{\delta x^{2} \delta t}} G_{i,m}
\end{equation}
where $G_{i,m}$ is taken from a zero-mean unit-variance Gaussian.

\subsection{Results from Bare Lattice Simulations }

Keeping the system always at the critical temperature $\theta_c$, we are
interested in its behavior as the free-energy barrier between the two
equilibrium phases is changed. We will measure the value of the
ensemble-averaged and
area-averaged order parameter $\langle\phi\rangle_A(t)\equiv 1/A\int d^2x
\phi({\bf x},t)$
for several choices of the
lattice spacing $\delta x$, taking note of its final equilibrium value,
${\bar \phi}_{\rm eq}$.
In figure 1 we show the results for $\langle\phi\rangle_A(t)$
for several choices
of lattice spacing and $\alpha = 0.45$.
The dependence on lattice-spacing
is quite
evident; different lattices produce different physics.

\begin{figure}
\psfig{figure=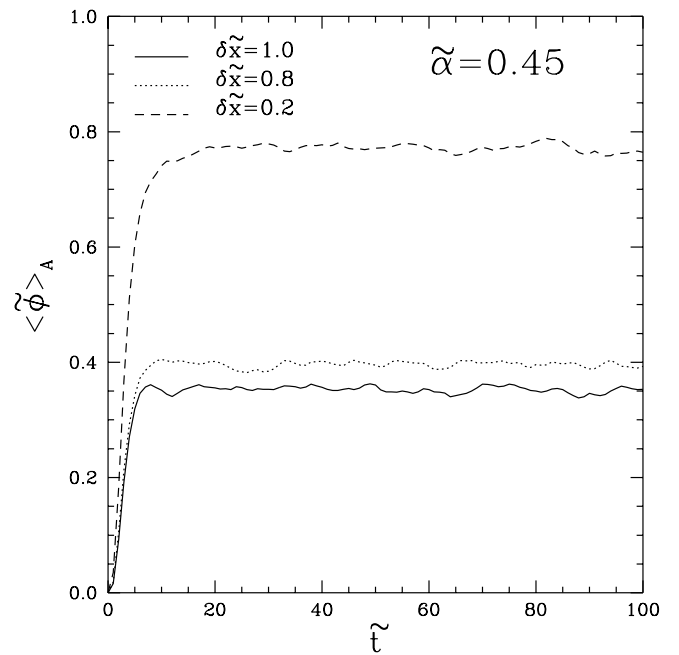,width=3.6in,height=3.6in}
\label{figure 1}
\caption{ $\langle\tilde{\phi}\rangle_A(\tilde{t})$ for the bare potential, for several choices
of lattice spacing, for $\tilde{\alpha} = 0.45$.}
\end{figure}

\begin{figure}
\psfig{figure=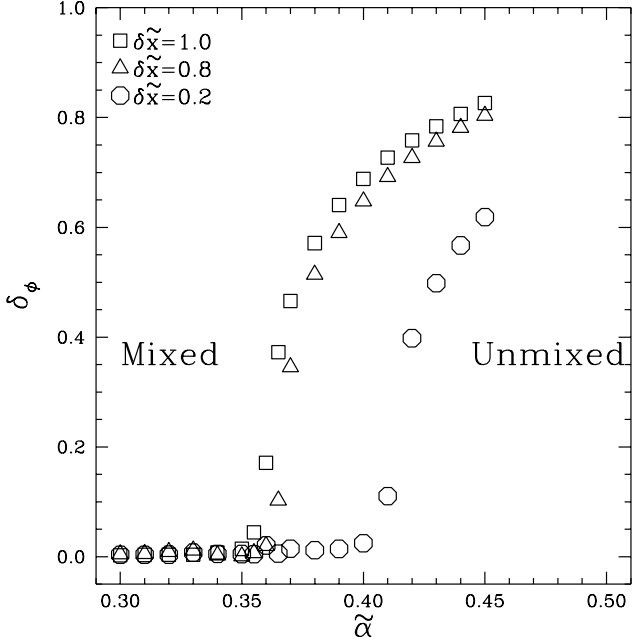,width=3.6in,height=3.6in}
\label{figure 2}
\caption{ Phase diagram for the bare potential for various lattice spacings. }
\end{figure}

In figure 2, we show the phase diagram depicting phase mixing as a function
of $\alpha$ for different choices of the lattice spacing $\delta x$. The
phase diagram is constructed by defining the ``phase-mixing order parameter'',
\begin{equation}
\delta_{\phi}(\alpha)\equiv |{\bar\phi}_{\rm eq}-
\phi_{\rm max}|/\phi_{\rm max}~~,
\label{phase mixing}
\end{equation}
where $\phi_{\rm max} =\alpha\theta_c/3\lambda$ is the
location of the maximum of
the free energy density separating the two phases.
Clearly, as $\alpha$ decreases, the free-energy barrier decreases and
larger-amplitude fluctuations between the two phases become more
probable. Below a critical value $\alpha_c$, ${\bar\phi}_{\rm eq}$ just
tracks the location of the maximum, indicating complete phase mixing, or
the breakdown of the mean field theory of Eq. \ref{hamiltonian}.

The problem, though, is that phase mixing, or the breakdown of mean-field 
theory,
occurs for values of $\alpha_c$
which are strongly dependent on the value of $\delta x$, as can be seen from
figure 2. For the range of
$\delta x$ investigated, $0.2\leq \delta x\leq 1$, we obtained $0.355 \lesssim
\alpha_c \lesssim 0.40$.
In the next
section, we argue that this dependence can be effectively cured by including
proper counterterms to the lattice potential.

\section{Approaching the Continuum on the Lattice}

\subsection{Computing the Lattice Effective Potential}

Setting up a continuum system on a lattice introduces two artificial
length scales, the ultraviolet momentum cutoff $\Lambda = \pi/\delta x$
and the infrared momentum cutoff $\Lambda_L=\pi/L$, where $L$ is the
lattice size. In the continuum limit, $L\rightarrow \infty$, and $\delta
x\rightarrow 0$ or, equivalently, the number of degrees of freedom
$N=(L/\delta x)^d \rightarrow \infty$. The coupling to the thermal bath
induces fluctuations at all allowed length scales. We should thus expect
that the lattice simulation is related to a continuum model with both
infrared and ultraviolet cutoffs. In order to obtain the lattice
effective potential, we start by analyzing the divergences of the
related continuous model.

For classical field theories in 2 dimensions, the corresponding 1-loop
corrected effective potential is given by
\begin{eqnarray}
V_{\rm 1L} (\phi) &=& V_0(\phi) + {T\over
2}\int_{\Lambda_L}^{\Lambda}{{d^2p}\over {(2\pi)^2}}
\ln\left (p^2 + V_0''\right ) \nonumber \\
\nonumber \\
& & \mbox{} +{\rm counterterms}~~,
\label{oneloop pot}
\end{eqnarray}
where the primes denote derivatives with respect to $\phi$. Performing
the integration and making all variables dimensionless we obtain,
\begin{eqnarray}
V_{\rm 1L}(\phi) &=& V_0(\phi) +
 \frac{\theta}{8 \pi} V_0''\left[
 1 - \ln{\left(\frac{\Lambda_L^2 + V_0''}{\Lambda^2} \right)} 
   \right]  \nonumber  \\
& & \mbox{} -  \frac{\theta}{8 \pi} \Lambda_L^2 \ln{\left(\Lambda_L^2 + V_0''\right)}
+ B \phi^2 \nonumber \\
\nonumber \\
& & \mbox{} + {\rm constants}~~,
\end{eqnarray}
The infrared cutoff does not introduce a divergence as
$\Lambda_L\rightarrow 0$, but it does introduce finite corrections to
$V_{\rm 1L}$, or finite size effects, which become small as $L$
increases. These become more severe near criticality, but well-known
scaling behavior can be used to regulate this dependence \cite{FINITESIZE}.
As we will further argue below, for our
purposes we can safely set $\Lambda_L=0$. This is not
the case for the ultraviolet cutoff. The reader can see now why it is
useful to use the shifted potential of Eq. \ref{shifted pot} as opposed
to the original one of Eq. \ref{gl pot}: all divergences are quadratic in
$\phi$, simplifying the computations considerably, while the physical
results, of course, remain unchanged. This is why we added only the
counterterm $B\phi^2$ above.

The counterterm $B$ is computed by imposing the renormalization condition
\begin{equation}
V_{1L}''(\phi_{\rm RN})=V_0''(\phi_{\rm RN})=M^2~~,
\end{equation}
where $M$ is the arbitrary renormalization scale and we write
$\phi_{\rm RN}\equiv \sqrt{\frac{M^2+\mu^2}{3 \lambda}}$.
[Note that $M$ here is
dimensionless (tilde is dropped), being defined as $\tilde M=M/T_2$.] One
obtains,
\begin{equation}
B(M)=\frac{\theta}{16 \pi} \left[V_0''''\ln{\left( \frac{V_0''}{\Lambda^2}
\right)} + \frac{(V_0''')^2}{V_0''} \right]_{\phi=\phi_{RN}}~~.
\label{counterterm}
\end{equation}

Applying this to the shifted potential of Eq. \ref{shifted pot}, we
obtain, for the 1-loop renormalized continuum potential,

\begin{eqnarray}
V_{1L}^M(\phi) &=& \left[ -\frac{1}{2}\mu^2+\frac{9 \lambda \theta}{8 \pi} +
\frac{3 \lambda \theta \mu^2}{4 \pi M^2}
 \right] \phi^2 + \frac{\lambda}{4}\phi^4  \nonumber   \\
& & \mbox{} + A \phi
 - \frac{3 \lambda \theta}{8 \pi} \phi^2
 \ln{\left( \frac{-\mu^2+3\lambda
\phi^2}{M^2} \right) }  \nonumber    \\
& & \mbox{} +  \frac{\mu^2 \theta}{8 \pi} \ln{(-\mu^2+3\lambda \phi^2)} +
{\rm constants}~~.
\label{1loop pot}
\end{eqnarray}

Recall that at $\theta_c$ the linear term proportional to $A(\theta)$ vanishes.
Since the counterterm cancels the dependence on the ultraviolet cutoff,
we define the lattice effective potential as \cite{BG}
\begin{equation}
V_{\rm latt}(\phi)=V_0(\phi) + B(M) \phi^2~~.
\label{lattice pot}
\end{equation}
In figure 3 we show the results of repeating the simulations of
figure 1 but now
adding the counterterm to the lattice simulations following Eq.
\ref{lattice pot}. The addition of the counterterm practically eliminates
the lattice-spacing dependence of the results.
Figure 3 also shows the near elimination of lattice-spacing dependence for $\alpha = 0.40$.

\begin{figure}
\psfig{figure=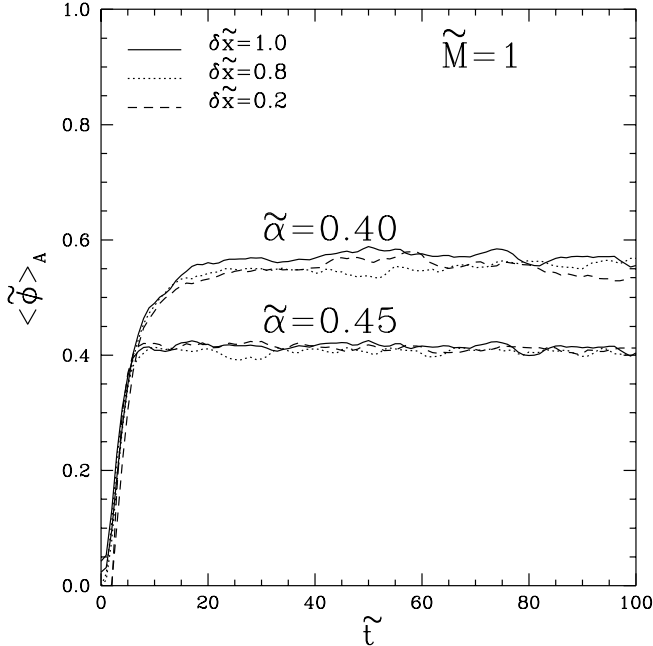,width=3.6in,height=3.6in}
\label{figure 3}
\caption{ ${\langle\tilde{\phi}\rangle_A}(\tilde{t})$
with the counterterm added and $\tilde{M}=1$, for several choices
of lattice spacing, for $\tilde{\alpha} = 0.45$ and $\tilde{\alpha} = 0.40$ }
\end{figure}

\subsection{Extracting the Critical Value of the Order Parameter}

In figures 4 and 5, we show the phase diagrams using $\delta_{\phi}$
defined in
Eq. \ref{phase mixing}
as a function of $\alpha$ for
different choices of lattice spacing $\delta x$. These are to be compared with
figure 2.
Figure 4 is for a choice of
renormalization scale $M=1$, while figure 5 is for $M=10$. It is
clear that the results for different lattice spacings converge around one
value of $\alpha_c$.

\begin{figure}
\psfig{figure=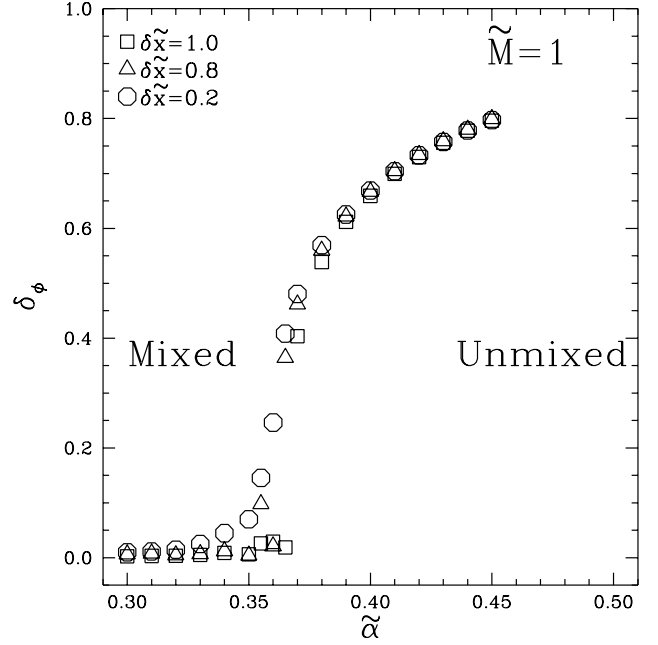,width=3.6in,height=3.6in}
\label{figure 4}
\caption{ Phase diagram for $\tilde{M}=1$.
}
\end{figure}

\begin{figure}
\psfig{figure=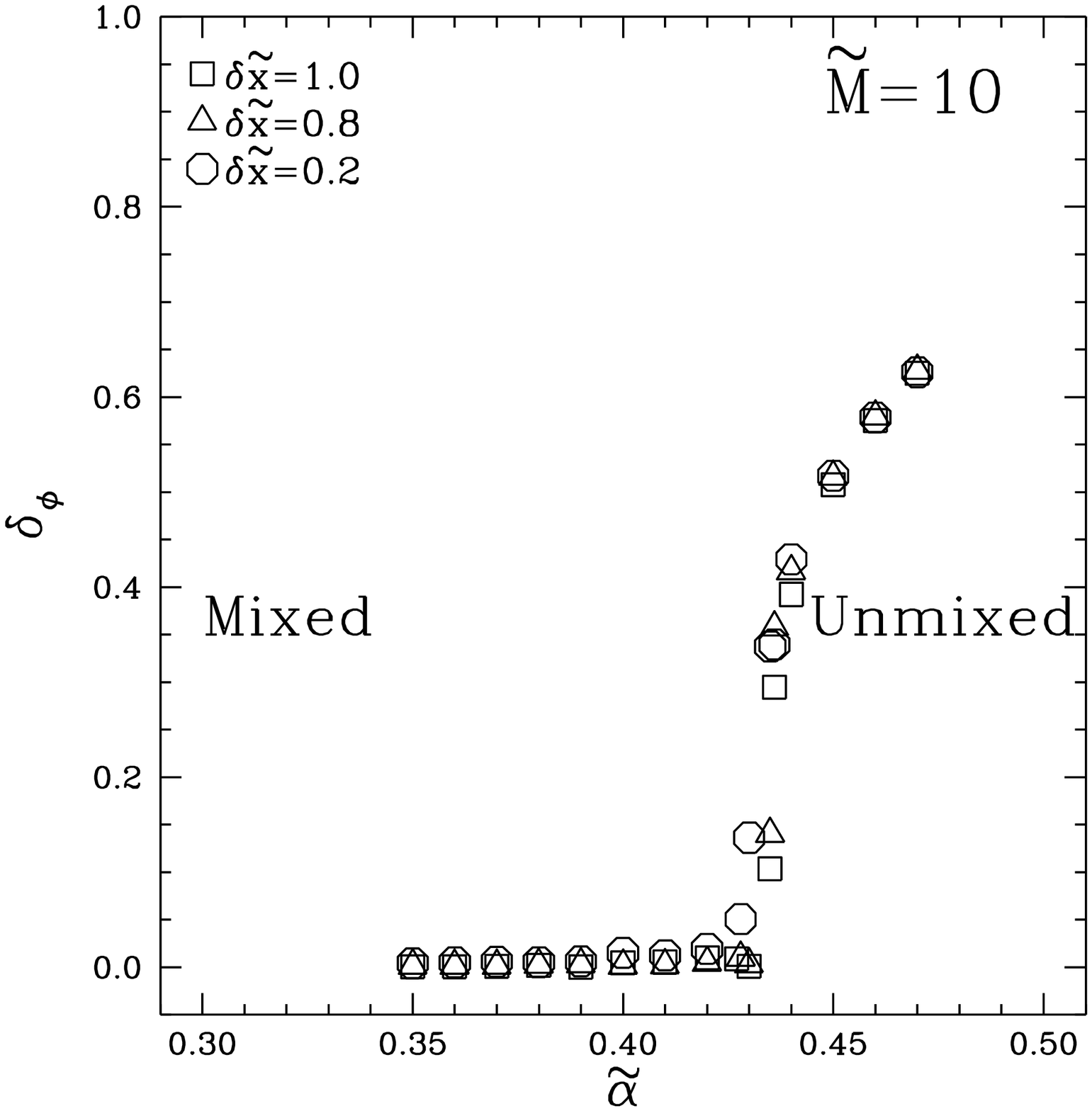,width=3.6in,height=3.6in}
\label{figure 5}
\caption{ Phase diagram for $\tilde{M}=10$.
}
\end{figure}

We compute
$\alpha_c$ as follows:
for a given value of $\alpha$ we perform several ($i_{\rm max}$)
measurements of ${\bar\phi}_{\rm eq}$ by varying the lattice spacing,
which we call ${\bar\phi}_{\rm eq}^i(\alpha)$. Their average is simply
$\langle {\bar\phi}_{\rm eq}(\alpha)\rangle =  \left [\sum_{1}^{i_{\rm max}}
{\bar\phi}_{\rm eq}^i(\alpha) \right ]/i_{\rm max}$,
while the departure from the average for each measurement is,
$\Delta {\bar\phi}_{\rm eq}^i =
|{\bar\phi}_{\rm eq}^i -\langle {\bar\phi}_{\rm eq}
\rangle |/\langle {\bar\phi}_{\rm eq}\rangle $.

Near criticality, the results are naturally poorer due to the existence of
long-range correlations in the field. We can use this fact to our advantage,
since we expect that, at criticality, the departure from the average defined
above is maximized, that is, the quantity
\begin{equation}
\langle \Delta {\phi}_{\rm eq} (\alpha) \rangle
\equiv { {\sum_1^{i_{\rm max}}\Delta {\bar\phi}_{\rm
eq}^i}\over {i_{\rm max}}}~~,
\label{alpha crit}
\end{equation}
reaches a maximum at $\alpha_c$. This can be clearly
seen from figure 6 for the same
choices
of lattice spacings (or coarse-graining scales) as in figures 4 and 5.

\begin{figure}
\psfig{figure=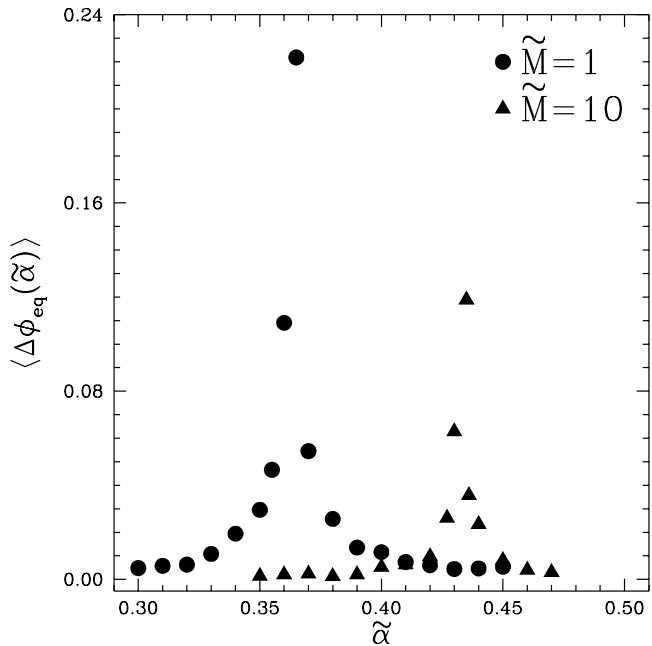,width=3.6in,height=3.6in}
\label{figure 6}
\caption{ $\langle \Delta {\phi}_{\rm eq} (\tilde{\alpha}) \rangle$ for $\tilde{M}=1$ and
$\tilde{M}=10$, with the respective
$\tilde{\alpha}_c$'s at the maxima.
}
\end{figure}

The
measured value of $\alpha_c$ is now $\alpha_c\simeq 0.365\pm 0.005$, for $M=1$,
and $\alpha_c\simeq 0.435\pm 0.005$ for $M=10$.

We have thus achieved lattice-spacing
independence on the measurement of $\alpha_c$. Clearly, the error in
$\alpha_c$ could be further decreased by taking a larger number of measurements
of ${\bar\phi}^i_{\rm eq}$. However, since our main goal
here is to show the convergence
of the results for different lattice spacings, we are not concerned with very
high-accuracy measurements. Nevertheless,
the values for $\alpha_c$ still depend on the renormalization scale,
which is arbitrary.
In the next subsection, we show how to obtain lattice results
which are independent of $M$.

\subsection{Achieving independence of renormalization scale on the
lattice}

As with conventional renormalization theory, the renormalized potential
should not depend on the choice of renormalization scale \cite{QFTTEXT}.
One usually solves the renormalization group equations to find how the
couplings vary with the scale. Here, we propose a simpler approach which
works quite well on the lattice implementation of scalar field theories.
It is an interesting question how to generalize it to more complex models.

Consider the 1-loop renormalized potential $V_{\rm 1L}^M(\phi)$ as given in
Eq. \ref{1loop pot}. The
superscript $M$ is a reminder that this potential is renormalized at a given
scale $M$. Now consider an equivalent potential renormalized at another
scale $M'$, $V_{\rm 1L}^{M'}(\phi)$. Since the divergences are quadratic,
this potential has a shifted mass $\mu'^2$. By imposing that the two potentials
are identical, $V_{\rm 1L}^M(\phi)=V_{\rm 1L}^{M'}(\phi)$, we obtain a
condition on the shifted mass $\mu'^2$, approximating
$\ln\left (-\mu'^2+3\lambda M'\right )\simeq
\ln\left (-\mu^2+3\lambda M\right )$,

\vspace{.15in}

\begin{equation}
{\mu'}^2\simeq \mu^2 + \frac{3 \lambda \theta}{4 \pi}
\ln\left ( \frac{M'^2}{M^2} \right )
 - \frac{3 \lambda \theta \mu^2}{2 \pi} \left [ \frac{1}{M^2} -
\frac{1}{M'^2}
\right ]~~.
\label{shifted mu}
\end{equation}
Thus, we can always relate a theory
with a choice of $M$ to any other theory with $M'$ by redefining the mass
$\mu^2$ according to Eq. \ref{shifted mu}. We claim that this is also the case
for the lattice effective potential.

As an illustration, we show the phase diagram for
$M'=10$ in figure 7, where the results for $M'=10$ were obtained after
scaling $\mu^2$ according to Eq. \ref{shifted mu} in the lattice potential
of Eq. \ref{lattice pot}.  It is practically indistinguishable from the
phase diagram
for  $M=1$ shown in figure 4.

\begin{figure}
\psfig{figure=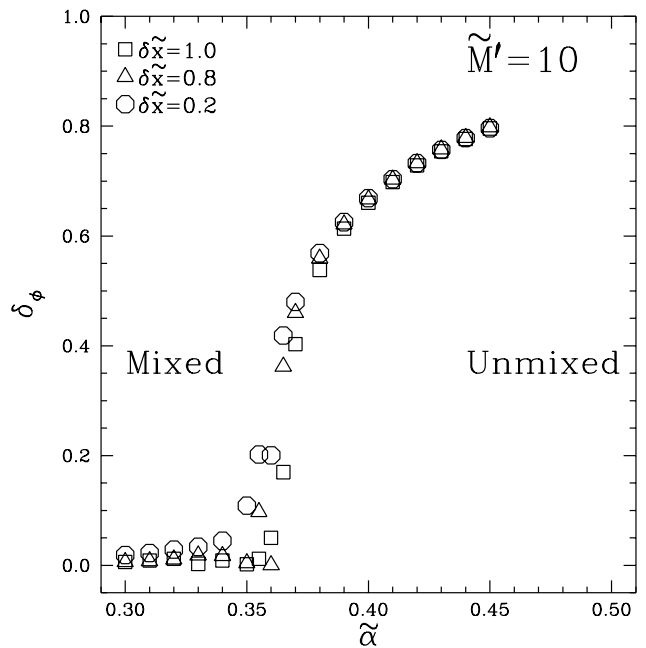,width=3.6in,height=3.6in}
\label{figure 7}
\caption{ Phase diagram for $\tilde{M}'=10$. }
\end{figure}

Figure 8 demonstrates clearly that $M'=10$ has the identical $\alpha_c$
previously found for $M=1$, within our level of accuracy.  This is in
stark contrast to figure 6, where
the values of $\alpha_c$ for $M=1$ and $M=10$ were very different, as evidenced
by its ``twin peaks'' structure.

\begin{figure}
\psfig{figure=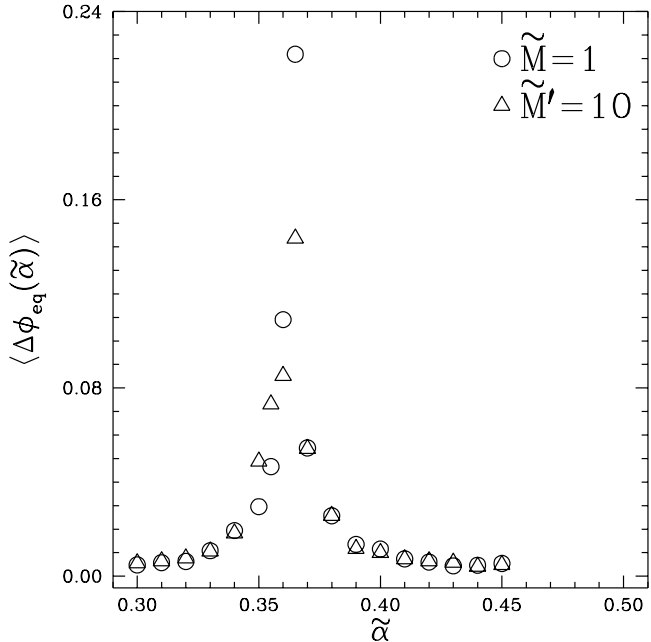,width=3.6in,height=3.6in}
\label{figure 8}
\caption{ $\langle \Delta {\phi}_{\rm eq} (\tilde{\alpha}) \rangle$ for $\tilde{M}=1$ and
$\tilde{M}'=10$ showing the same value
 of $\tilde{\alpha}_c$.
}
\end{figure}

\section{Summary and Outlook}

We have investigated the continuum limit of lattice simulations of
stochastic scalar field theories. In particular, we have proposed a method
to obtain not only lattice-spacing independent results, but also results
independent of the renormalization scale of the lattice effective potential.
We illustrated our approach by examining a Ginzburg-Landau model
which exhibits phase mixing depending on the values of the parameters
controlling the free-energy barrier for large-amplitude fluctuations between
the two low-temperature phases in our model. Thermal fluctuations of the
order parameter are induced
by coupling it to a thermal bath at fixed temperature $T_c$, defined
as the temperature where the two phases have the same free energy density.
We simulate the dynamics using a generalized Langevin equation with
Gaussian noise, which brings the system to its final equilibrium state.

The results were presented in terms of phase diagrams which clearly
illustrate the effectiveness of our approach. We also proposed a simple way
of determining the critical value of the control parameter for phase mixing,
which uses the spread in values of the equilibrium order parameter
around criticality for different choices of lattice spacing
(or coarse-graining scales). Thus, we effectively turn a
weakness of lattice simulations
into a strength, something that can be useful for the examination of
critical phenomena of continuous field theories in fairly small lattices.

We plan to expand the present study to investigate the effects of
spatio-temporal memory on the dynamics of nonequilibrium fields. Recent
results have shown that the effective Langevin equation for self-coupled
scalar systems exhibits colored and multiplicative noise \cite{GR}. It is
possible to expand the two-point function characterizing the noise (or noises)
in terms of a ``persistence factor'', which defines short or long-term memory,
spatial, temporal, or both. The possible impact of this kind of
noise on the nonequilibrium dynamics of fields remains largely unexplored.

\section{Acknowledgements}

C.G. was supported in part by NASA through the New Hampshire
NASA Space Grant (NGT5-40010).
M.G. was supported in part
by an NSF Presidential Faculty Fellows award PHY-9453431. M.G. thanks
the Particle Theory Group at Boston University, where parts of this work
were developed, for their hospitality.

Carmen Gagne: carmen.gagne@dartmouth.edu

Marcelo Gleiser: marcelo.gleiser@dartmouth.edu

\end{document}